\documentclass[aps,twocolumn,showpacs,preprintnumbers,amsmath,amssymb,pre]{iopart}
\usepackage{epsf, iopams, amssymb,verbatim,color,multirow,pifont}
\usepackage{graphicx}

\begin{document}
\title{Discontinuous percolation in diffusion-limited cluster aggregation}
\author{Y.S. Cho$^1$, Y. W. Kim$^{1,2}$ and B. Kahng$^1$}
\address{{$^1$Department of Physics and Astronomy, Seoul National University,
Seoul 151-747, Korea}\\
{$^2$ Department of Physics, Lehigh University, Bethlehem,  PA 18105, USA} }
\ead{koreafire@hanmail.net, ywk0@lehigh.edu and bkahng@snu.ac.kr}
\date{\today}

\begin{abstract}
Recently, the diffusion-limited cluster aggregation (DLCA) model was restudied as a real-world example of showing discontinuous percolation transitions (PTs). Because a larger cluster is less mobile in Brownian motion, it comes into contact with other clusters less frequently. Thus, the formation of a giant cluster is suppressed in the DLCA process. All clusters grow continuously with respect to time, but the largest cluster grows drastically 
with respect to the number of cluster merging events. Here, we study the discontinuous PT occurring in the DLCA model in more general dimensions such as two, three, and four dimensions. PTs are also studied for a generalized velocity, which scales with cluster size $s$ as $v_{s} \propto s^{\eta}$. For Brownian motion of hard spheres in three dimensions, the mean relative speed scales as $s^{-1/2}$ and the collision rate $\sigma v_s$ scales as $\sim s^{1/6}$. We find numerically that the PT type changes from discontinuous to continuous as $\eta$ crosses over a tricritical point $\eta_{c} \approx 1.2$ (in two dimensions), $\eta_{c} \approx 0.8$ (in three dimensions), and $\eta_{c} \approx 0.4$ (in four dimensions). We illustrate the root of this crossover behavior from the perspective of the heterogeneity of cluster-size distribution. Finally, we study the reaction-limited cluster aggregation (RLCA) model in the Brownian process, in which cluster merging takes place with finite probability $r$. We find that the PTs in two and three dimensions are discontinuous even for small $r$ such as $r=10^{-3}$, but are continuous in four dimensions.   
\end{abstract}

 \maketitle


\section{Introduction}
The notion of percolation is widely used to explain the formation of a macroscopic spanning cluster in diverse 
systems \cite{stauffer}. In percolation, as the control parameter, i.e., the density of occupied nodes (in site percolation) or 
bonds (in bond percolation), is increased, a macroscopic spanning cluster emerges at the percolation threshold. 
This behavior is referred to as the percolation transition (PT), which is conventionally continuous. 
More generally, the term PT is used for the emergence of a macroscopic cluster in growing networks, and is occasionally referred to as PT. In the random graph model introduced by Erd\H{o}s and R\'enyi (ER) \cite{er}, the control parameter is the number of bonds (links), and a macroscopic giant cluster is found to emerge at a critical point.  
That is, PT occurs at a finite percolation threshold. Recently, Achlioptas {\it et al.} \cite{ap} proposed a modified 
ER model,  in which the growth of the largest cluster is suppressed during the dynamical evolution.
In this model, the macroscopic giant cluster emerges at a delayed transition point and the transition occurs in a rather 
explosive manner. Thus, the PT in this model is claimed to be discontinuous. Following the proposal of this explosive percolation model, many studies have been performed on discontinuous PT; however, whether such explosive percolation transitions are indeed discontinuous in the thermodynamic limit is still a matter of debate and sensitive to detailed dynamic rules 
\cite{friedman, fortunato, dorogovtsev, hklee, grassberger, science, suppression, souza, hermann, choi, ziff}.
Nonetheless, the introduction of such explosive percolation models has led to intensive studies of discontinuous PT. In this circumstance, we wonder whether such discontinuous PTs indeed can be observed in real-world systems. 

In our previous work \cite{dlca}, we studied the diffusion-limited cluster aggregation (DLCA) model following Brownian motion in two dimensions for simplicity, as an example of a real-world system showing discontinuous PT. 
Here, the PT means the formation of a giant component, as conventionally used in the evolution of random graphs, 
instead of the formation of a spanning cluster, as conventionally used in regular lattice.
We monitored the PT as a function of the number of cluster aggregations. Because real-world systems can be three dimensional, here we extend our previous study to three- and four-dimensional cases. Moreover, we study the PT for the case of cluster velocity being in the general form $v_{s} \propto s^{\eta}$ and find that the PT type changes from discontinuous to continuous as $\eta$ increases. In the last part of this paper, we extend our study to the reaction-limited cluster aggregation (RLCA) model in which clusters diffuse following Brownian motion, and when two clusters come into contact with each other, they merge with a certain probability $r$ and remain separate with the remaining probability $1-r$. We find that the discontinuous PT behavior 
can also be observed in this RLCA model in two and three dimensions but that the PT remains continuous in four dimensions. 
 
\section{Diffusion-limited cluster aggregation model}

The DLCA model was introduced by Meakin {\it et al.} \cite{meakin} and Kolb {\it et al.} \cite{kolb}. Initially, $N$ particles are placed randomly in a $d$-dimensional lattice space with linear size $L$. The density of the particles is fixed as $\rho=N/L^d$, whereas the system size $L$ is controllable. Simulations start from $N$ monoparticles.  
When an $s$-sized cluster moves in the Brownian process, its velocity is given as $v_s \sim s^{-1/2}$, and 
the collision rate per cluster becomes $\sigma v_s \sim s^{1/6}$ in three dimensions when the cluster is 
regarded as a hard sphere \cite{kim}. This originates from the fact that a Brownian particle with mass $m$ has 
mean velocity $\overline{v}=0$ and velocity fluctuations $\overline{v^2} \sim k_BT/m$ when the particle is in 
thermal equilibrium with temperature $T$, where the overbar means ensemble average over thermal fluctuations 
and $k_B$ is the Boltzmann constant. Thus, we obtain $\sqrt{\overline{v^2}}\sim 1/\sqrt{m}$, 
which leads to $v_s \sim s^{-1/2}$ when cluster mass is regarded as being linearly proportional to cluster size. 
Accordingly, the choice of a cluster of size $s$ with probability $s^{-1/2}/\sum_s N_s s^{-1/2}$ leads to 
$v_s \sim s^{-1/2}$. 

To implement this velocity in simulations, we perform a simulation in the following steps: Initially, all particles are single. (i) An $s$-sized cluster is picked up with a probability proportional to $s^{-1/2}$, and it is moved to one of the nearest-neighbor positions. All particles in a mobile cluster move together with the cluster shape unchanged. After this move, when two distinct clusters come into contact, those clusters merge with probability one, forming a larger cluster. (ii) Time is advanced by $\delta t=1/(\sum_s N_s s^{-1/2})$, whenever the cluster moves irrespective of whether a contact occurs, but the control parameter $p$ is advanced by $1/N$ only when the cluster is placed next to another cluster and the two clusters merge. When all particles merge into a single cluster, the dynamics ends. 

We presume that cluster aggregations take place irreversibly. Then the number of cluster merging events during the whole process is $N-1$. For example, if two immobile clusters are merged by one mobile cluster, and thus the three clusters become one, then the number of merging events is counted as two, and $p$ is advanced by $2/N$. Thus, the number of merging events corresponds to the number of inter-cluster edges connected. When a created cluster contains loop structure, then the number of cluster merging events is not same as the number of occupied bonds in bond percolations.

The parameter $p$ represents the number of cluster merging events per total particle number, and corresponds to the number of links connecting two distinct clusters per network size in the random graph model. The variable $p$ turns out to differ from time $t$ in a nontrivial way. The order parameter of the PT is the giant cluster size per system size $N$, denoted as $G_N \equiv G$. To examine the PT, $G_N$ is measured as a function of $p$, which becomes nonzero beyond the transition point $p_c$, where PT is discontinuous if $G_N(p_c^+)-G_N(p_c^-) > 0$ and continuous if it is zero.    

\subsection{Brownian motion}

\begin{figure}
\includegraphics[width=0.8\linewidth]{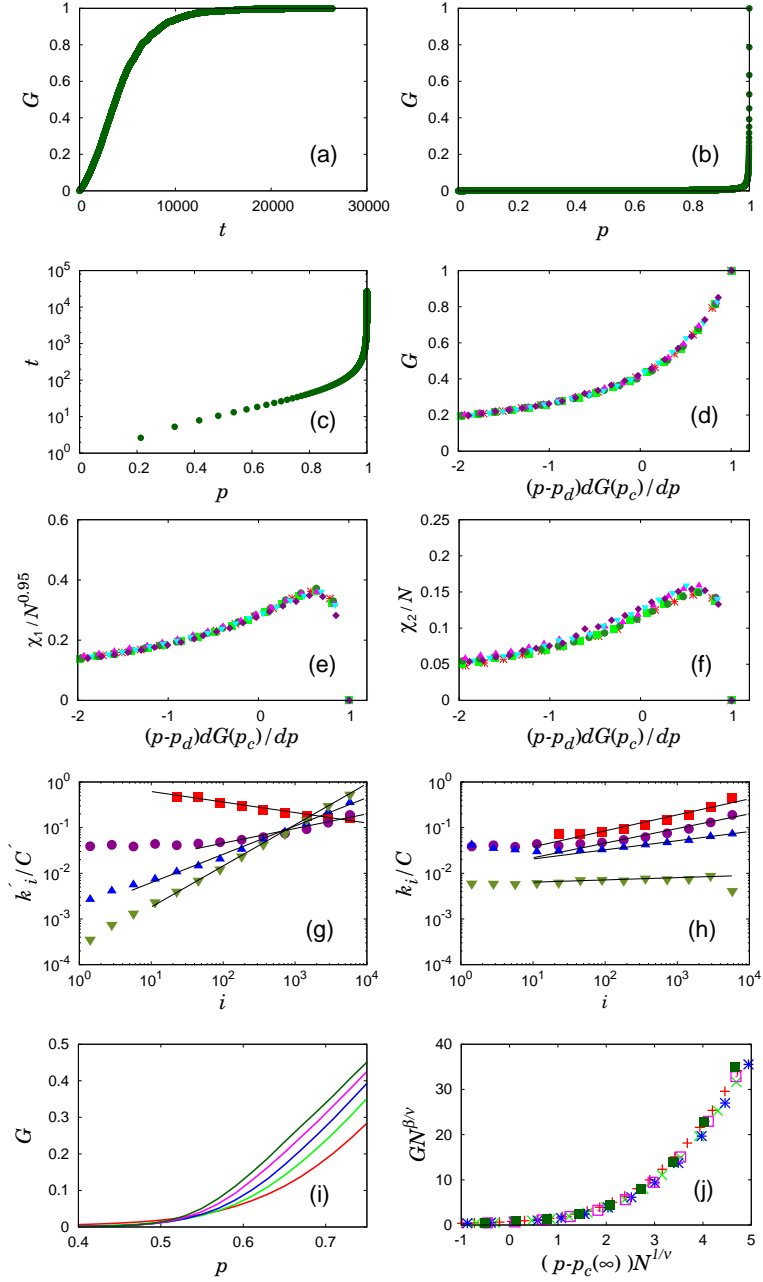}
\caption{(Color online) (a)-(f) Simulation results of the DLCA model in two dimensions with velocity 
exponent $\eta = -0.5$ for Brownian motion. (a) Plot of $G$ vs $t$. 
$G$ grows monotonically from $t = 0$ with increasing $t$. (b) $G$ grows discontinuously near $p \approx 1$ with respect to $p$. (c) Relation between $t$ and $p$. $t$ grows rapidly as $p$ approaches $p=1$, by which a discontinuous PT occurs with respect to $p$. (d)-(f) Finite-size scaling analysis for the discontinuous PT of $G$, $\chi_1$, and $\chi_2$ using Eqs.~(\ref{g_scaling}-\ref{susc2}). Data of different-size systems collapse well onto a single curve, which implies that a discontinuous PT occurs. We use the system sizes $L/10^2 = 6, 10, 14, 18, 22$, and $26$ with density $\rho = 0.05$ for numerical simulations.
(g) Numerical estimations of $k^{\prime}_i/C^{\prime}$ at $p_d$ for $\eta = -0.5 (\square)$, 
$\eta = 0 (\diamond)$, $\eta = 0.4 (\triangle)$, and $\eta = 0.8 (\nabla)$. Simulations are performed in systems with $N = 8000$ and $L = 400$. Slopes of the guidelines are $-0.23 \pm 0.02$, $0.32 \pm 0.04$, $0.62 \pm 0.01$, and $0.88 \pm 0.01$ from above.    
(h) Numerical estimations of  $k_i/C$ with the same symbols as in (g) used for each $\eta$. Here $N = 8000$ and $L = 400$ are used for simulations. Slopes of the guidelines are $0.35 \pm 0.04$, $0.32 \pm 0.04$, $0.2 \pm 0.01$, and $0.05 \pm 0.01$ from above.
(i) Plot of $G$ vs $p$ of different size systems with $\eta = 1.5$ in two dimensions. A giant component emerges continuously near $p \approx 0.45$.
Data of $N/10^3 = 8, 32, 72, 128$, and $200$ with $\rho = 0.05$ are shown in this plot.
(j) Finite-size scaling analysis for continuous PT for the data used in plot (i). $p_c(\infty) = 0.43$, $1/\nu = 0.29$, and $\beta/\nu = 0.5$ are estimated.}\label{fig1}
\end{figure}

We begin by recalling our previous work in two dimensions. The giant cluster size increases monotonically as a function of time $t$; however, it increases drastically when monitored as a function of the variable $p$ as shown in Figs.\ref{fig1}(a) and 1(b), respectively. This indicates that the difference originates from the nonlinear relationship between $t$ and $p$ shown in Fig.\ref{fig1}(c). In particular, the time interval between two successive cluster merging events becomes long when $p$ approaches one, because few clusters remain and they hardly ever contact each other. Thus, such a nontrivial relationship between $t$ and $p$ arises. 

To verify the discontinuity of the order parameter, we use the finite-size scaling approach, which is different from 
the conventional one used for the continuous PT \cite{fss}.  
In this approach, a particular point $p_d(N)$ was introduced as a triggering
\begin{equation}   
p_d(N)=p_c(N)-\Big(\frac{dG_N(p)}{dp}\Big|_{p_c}\Big)^{-1}G_N (p_c), \label{eq1}
\end{equation}
where $p_c(N)$ is the point at which the slope of $G_N(p)$ becomes maximum.
It is found that $dG_N(p)/dp |_{p_c}$ increases in a power-law manner $\sim N^{1/\overline{\nu}}$ 
with $1/\overline{\nu} \approx 0.86 \pm 0.02$. Then, since the giant component size
grows as $G_N(p_f)-G_N(p_d) \sim \mathcal{O}(1)$, where $p_f$ is the final step of cluster aggregation, 
during the interval $p_f-p_d \sim \mathcal{O}(N^{-1/\overline{\nu}})$, the transition is indeed discontinuous. 
The above behavior is also checked by using the scaling ansatz  for the discontinuous PT,
\begin{equation}
G_N(p)\propto N^{-\beta/{\bar \nu}}f_0((p-p_d(N))N^{1/{\bar \nu}}),\label{g_scaling}
\end{equation}
where $\beta/\overline{\nu}=0$ for the discontinuous transition.
This scaling form differs from that conventionally used for continuous transitions, which is written as 
Eq.~(\ref{cont_fss}) shown later.
Thus a discontinuous PT can be confirmed by checking whether the data of $G_N(p)$ versus $f_0(\overline p)$ with 
$\overline{p}=(p-p_d)N^{1/\overline{\nu}}$ for different system sizes collapse onto a single curve or not. 
Indeed, we find that the data from different system sizes collapse onto a single curve  
in two dimensions when using the value $1/\overline{\nu}=0.86$ in Fig.~\ref{fig1}(d).

\begin{figure}
\includegraphics[width=0.9\linewidth]{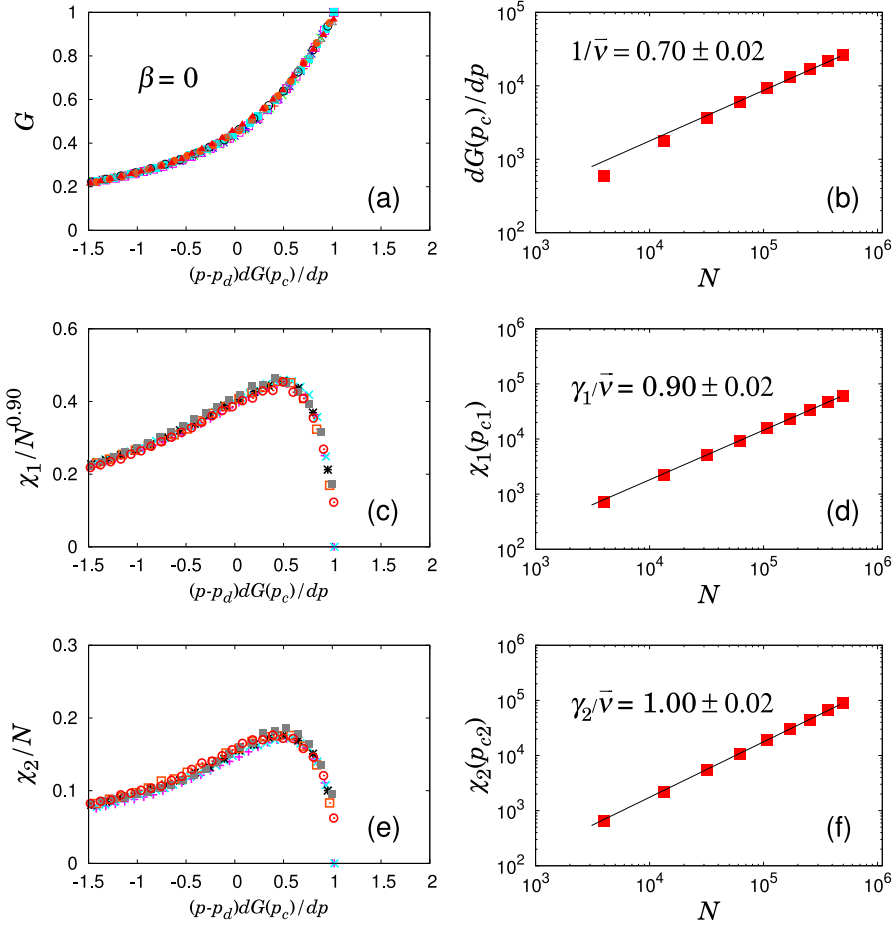}
\caption{(Color online) Finite-size scaling analysis for discontinuous PT of the DLCA model in the Brownian process $(\eta = -0.5)$ in three dimensions having linear sizes $L/10 = 25, 30, 35, 40, 
45$, and $50$ with $\rho = 0.004$.
(a) Data collapse of $G$ by using Eq.~(\ref{g_scaling}) for different size systems.
(b) The estimation of the exponent $1/\overline{\nu}$ defined in the relation 
$dG(p)/dp |_{p_c} \sim N^{1/\overline{\nu}}$. We obtain $1/\overline{\nu}=0.70\pm0.02$. 
(c) Plot of $\chi_1$ for different system sizes in a scaling form. 
(d) The estimation of the exponent $\gamma_{1}/\overline{\nu}$ defined in the relation $\chi_{1}(p_{c1}) \sim N^{\gamma_{1}/\overline{\nu}}$. We obtain $\gamma_{1}/\overline{\nu} = 0.90 \pm 0.02$. (e) Data collapse of $\chi_{2}$ of different system sizes in a scaling form. 
(f) The estimation of the exponent $\gamma_{2}/\overline{\nu}$ defined in the relation $\chi_{2}(p_{c2}) \sim N^{\gamma_{2}/\overline{\nu}}$. We obtain $\gamma_{2}/\overline{\nu}=1.00 \pm 0.02$.
In particular, we can find that the peak height of the standard deviation of the largest cluster increases proportionally to the system size, which is a feature of the first-order transition.}\label{fig2}
\end{figure}

\begin{figure}
\includegraphics[width=0.9\linewidth]{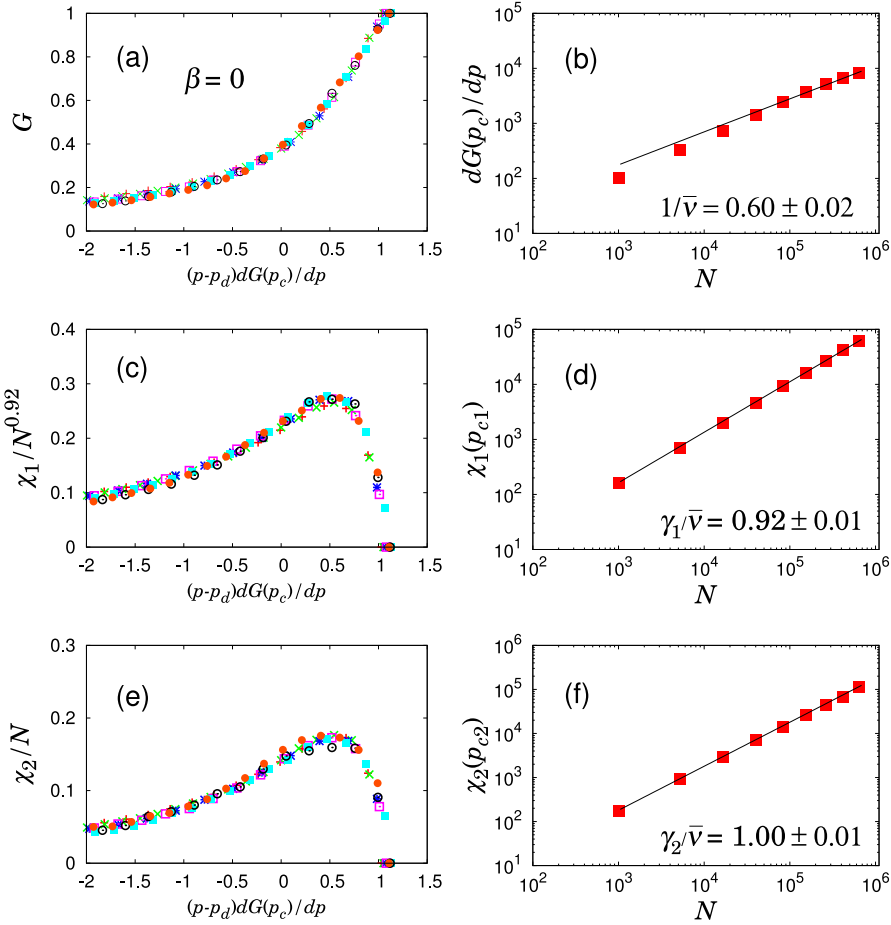}
\caption{(Color online) Finite-size scaling analysis for discontinuous PT of the DLCA model in Brownian motion $(\eta = -0.5)$ in four dimensions. 
We use the scaling forms of Eq.~(\ref{g_scaling}) for $G$ in (a), Eq.~(\ref{susc1}) for $\chi_1$ in (c), and Eq.~\ref{susc2} for $\chi_2$ in (e). Here, the systems having linear size 
$L/10 = 8, 10, 12, 14, 16, 18$, and $20$ with $\rho = 0.0004$ are used for simulations. 
These data are well collapsed in each scaling form, which confirms that the PT is indeed a discontinuous PT. (b), (d), and (f) Numerical estimations of the exponents used in each scaling form. In particular, we can find $\gamma_2/\overline{\nu}=1$, which is a feature of the first-order transition.}
\label{fig3}
\end{figure}

\begin{figure}
\includegraphics[width=0.9\linewidth]{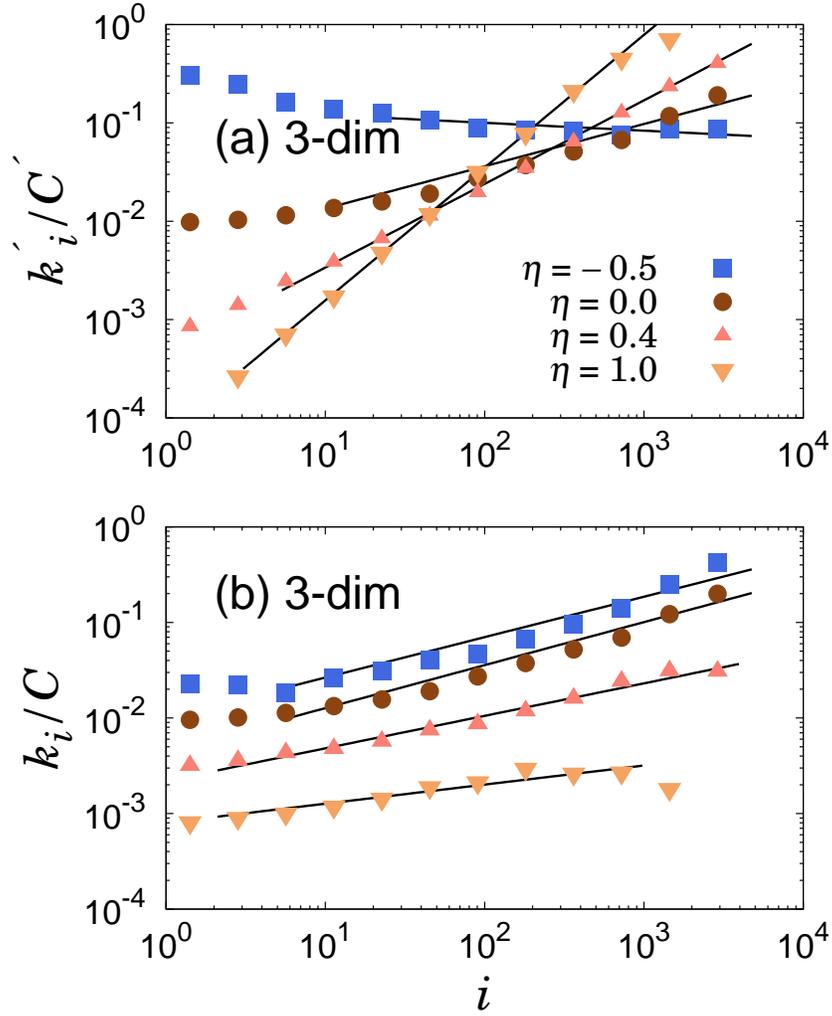}
\caption{(Color online) Plots of (a) $k^{\prime}_i/C^{\prime}$ and (b) $k_i/C$ vs $i$ at $p_d$ of the DLCA model in three dimensions. The system of $N=4000$ and $L=100$ is used. (a) The estimated slopes are $-0.08 \pm 0.03$ $( \eta = -0.5 )$ $(\square)$, $0.43 \pm 0.02$ $(\eta = 0.0)$ $(\circ)$, $0.85 \pm 0.01$ $(\eta = 0.4)$ $(\triangle)$, and $1.35 \pm 0.01$ $(\eta = 1.0 )$ $(\triangledown)$. 
(b) The estimated slopes are $0.43 \pm 0.03$ $(\square)$, $ 0.45 \pm 0.03$ $(\circ)$, $0.34 \pm 0.03$ $(\triangle)$, and $0.20 \pm 0.03$ $(\triangledown)$.}\label{fig4}
\end{figure}

\begin{figure}
\includegraphics[width=0.9\linewidth]{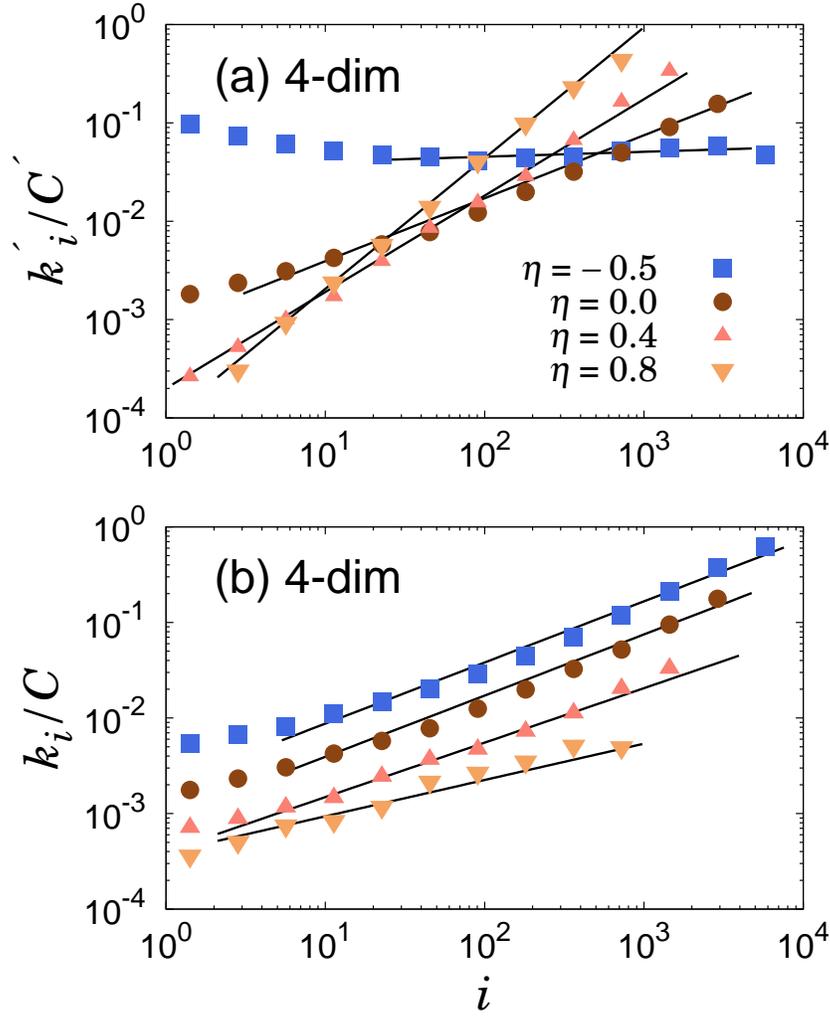}
\caption{(Color online) Plots of (a) $k^{\prime}_i/C^{\prime}$ and (b) $k_i/C$ vs $i$ at $p_d$ of the DLCA model in four dimensions. Simulations are carried out in the system of $N=5184$ and $L=60$.
(a) The estimated slopes are $0.04 \pm 0.02 $ $(\eta = -0.5)$ $(\square)$, $0.64 \pm 0.02 $ $(\eta = 0.0)$ $(\circ)$, $0.97 \pm 0.04$ $(\eta = 0.4)$ $(\triangle)$, and $1.33 \pm 0.02$ $(\eta = 0.8)$ $(\triangledown)$ from above. 
(b) The estimated slopes are $0.64 \pm 0.03 (\square)$, $0.64 \pm 0.02 (\circ)$, $0.57 \pm 0.02 (\triangle)$, and $0.38 \pm 0.04 (\triangledown)$.}
\label{fig5}
\end{figure}

\begin{figure*}
\includegraphics[width=0.9\linewidth]{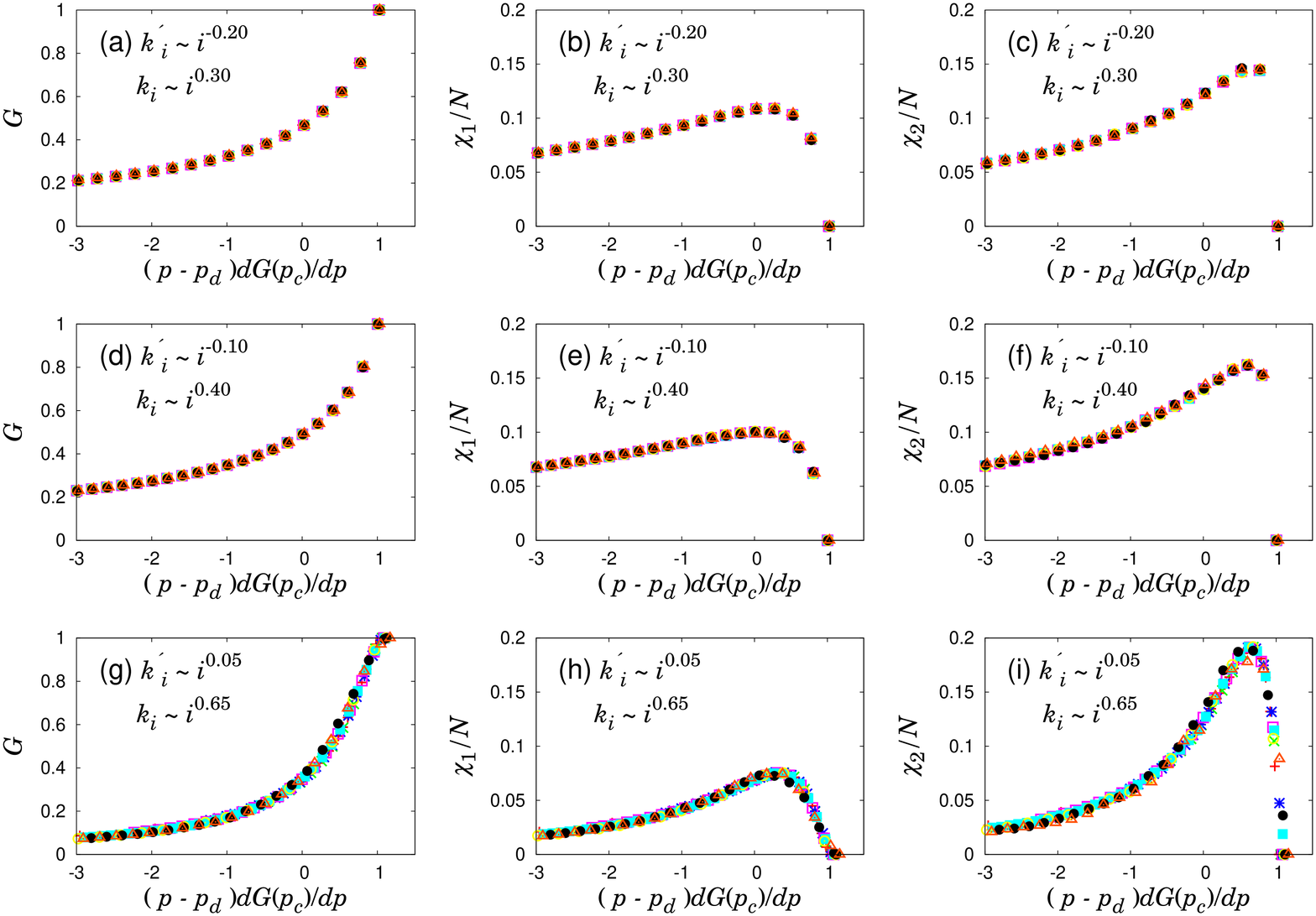}
\caption{(Color online) Finite-size scaling analysis for the discontinuous PT based on the numerical simulation results of the Smoluchowski equation. Collision kernels obtained from the data in Fig.~\ref{fig1}, Fig.~\ref{fig4}, and Fig.~\ref{fig5} for the Brownian case are used in two, three, and four 
dimensions, respectively. The system sizes we used are $N/10^4 = 16, 32, 64, 128, 256, 512, 1024$, 
and $2048$. (a)-(c) In two dimensions, $dG(p_c)/dp$ behaves as $\sim N$. By using this result, data collapse 
behaviors are obtained for (a) the giant cluster size, and the susceptibilities (b) $\chi_1$ and (c) $\chi_2$.
(d)-(f) Similar analyses are carried out for three dimensions, in which $dG(p_c)/dp \sim N$.
(g)-(i) Similar analyses are carried out for four dimensions, in which $dG(p_c)/dp \sim N^{0.54}$.}
\label{fig6}
\end{figure*}

\begin{figure}
\includegraphics[width=0.9\linewidth]{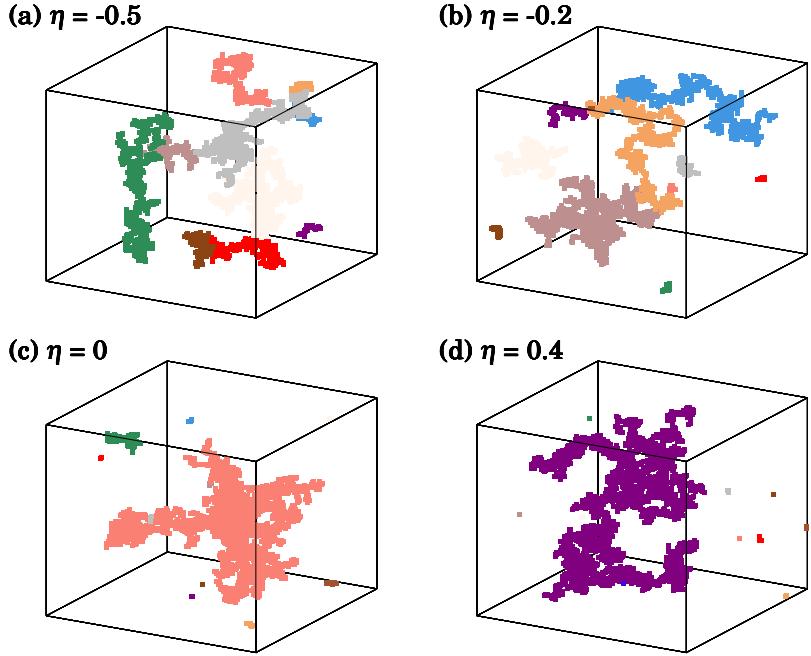}
\caption{(Color online) Snapshot of clusters in a three-dimensional lattice at $p = 0.997$ with 
the system size $N = 4000$ and $L = 100$. There are eleven clusters for each snapshot. 
We find that cluster sizes are more heterogeneous as $\eta$ grows. This means that the 
rate of growth of the largest cluster increases as $\eta$ increases.}\label{fig7}
\end{figure}

\begin{figure}
\includegraphics[width=0.9\linewidth]{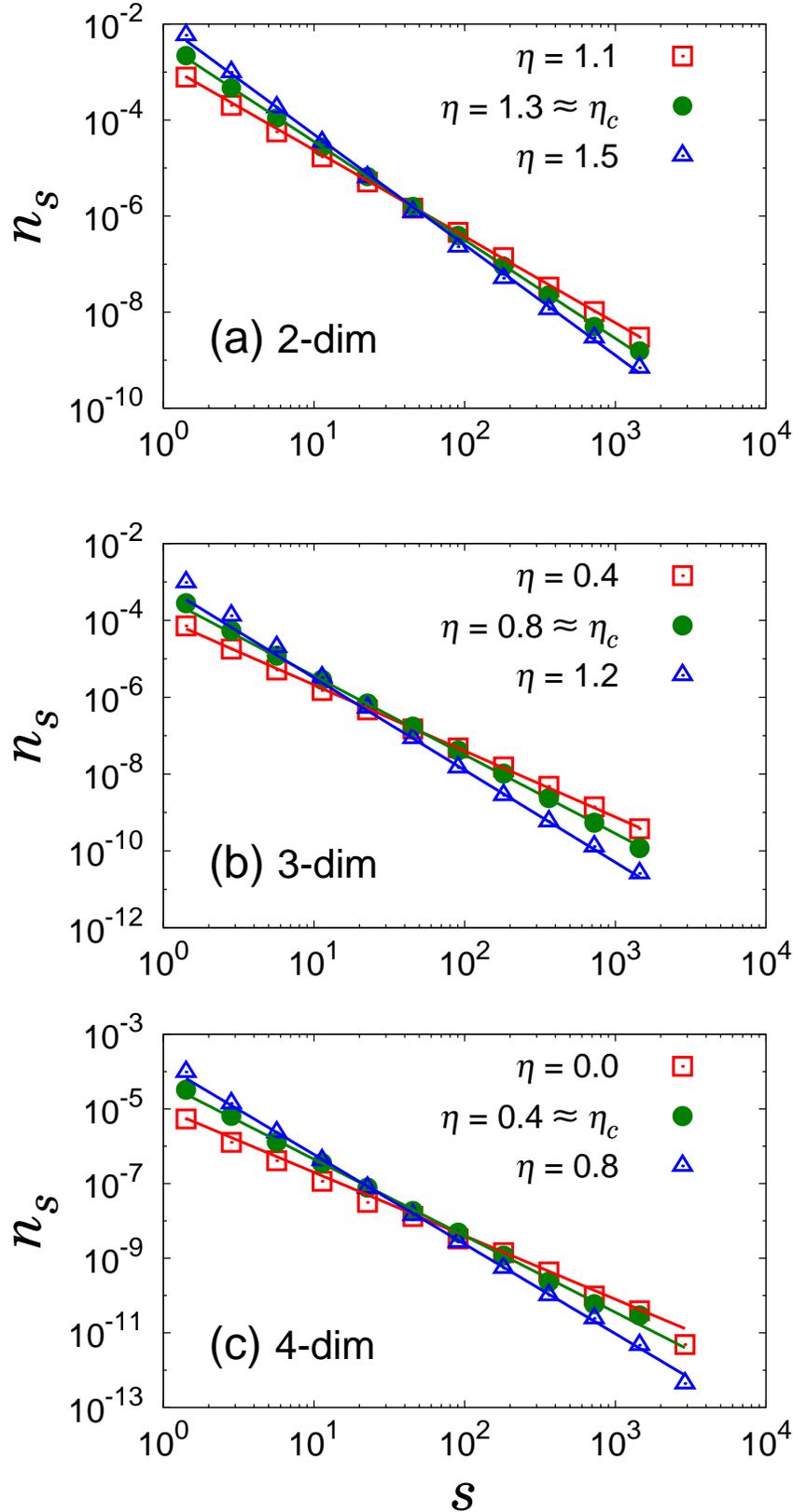}
\caption{(Color online) Plots of the cluster size distribution $n_s(p_c^+) \sim s^{-\tau}$ 
for several $\eta$ around the tricritical point $\eta_c$.
The slopes of the guidelines are (a) $-1.8$, $-2.0$, and $-2.3$ in two dimensions, 
(b) $-1.7$, $-2.0$, and $-2.4$ in three dimensions, and (c) $-1.7$, $-2.0$, and $-2.4$ in four 
dimensions. The system sizes we used are (a) $N=8,000$, $L=400$, (b) $N = 13,500$, $L = 150$, and (c) $N = 40,000$, $L = 100$.}\label{fig8}
\end{figure}

\begin{figure}
\includegraphics[width=0.9\linewidth]{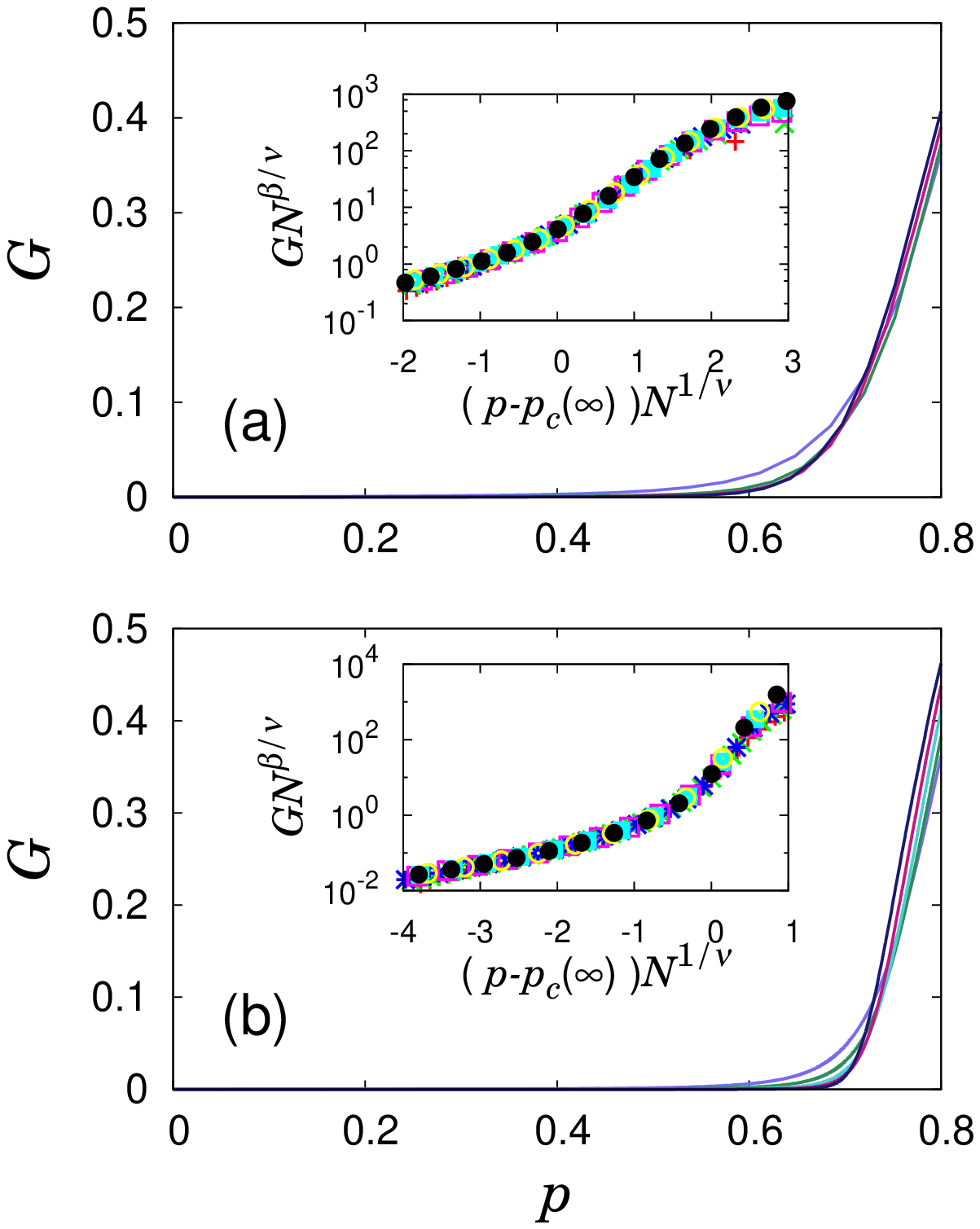}
\caption{(Color online) (a) Plot of the giant cluster size $G$ vs $p$ of the DLCA model 
with velocity exponent $\eta=1.0 > \eta_c \approx 0.8$ in three dimensions. Data are obtained 
by Monte Carlo simulations. (b) Numerical simulation of the Smoluchowski equation. 
In (b), we use the collision kernel $k^{\prime}_i/C^{\prime} \sim i^{1.35}$ and 
$k_i/C \sim i^{0.20}$.  Inset in (a): Data collapse of the data plotted in (a). Linear sizes of the systems are $L/10 = 10, 15, 20, 25, 30, 35$, and $40$ and the density of particles is $\rho = 0.004$. 
Data of different system sizes collapse well onto a single curve predicted theoretically for a continuous transition in which we use $p_c(\infty)=0.56$, $1/\nu=0.2$, and $\beta/\nu=0.6$. The behavior of the data collapse indicates 
that the transition is indeed continuous. Inset in (b): Data collapse of the data plotted in (b). We use system sizes 
 $N/10^4 = 8, 32, 128, 512$, and $2048$ and substitute $p_c(\infty)=0.66$, $1/\nu=0.14$, and 
 $\beta/\nu=0.55$ for data collapse. Again the data-collapse behavior indicates a continuous PT.}
\label{fig9}
\end{figure}

We also studied the susceptibility, defined as $\chi_1(p)\equiv \sum_s^{\prime} s^2 n_s(p)/\sum_s^{\prime} sn_s(p)$, 
where the prime represents the exclusion of the giant component in summation.  This function can be 
represented in the scaling form, 
\begin{equation}
\chi_1(p)\propto N^{\gamma_1/\overline{\nu}}f_1((p-p_d)N^{1/\overline{\nu}}).\label{susc1}
\end{equation}
It was found that the data from different system sizes collapsed well onto a single curve with 
the exponent value $\gamma_1/\overline{\nu}=0.95$ in Fig. \ref{fig1}(e).

We also attempt a scaling analysis for another quantity of the susceptibility $\chi_2$ defined as
$\chi_2 \equiv N\sqrt{\langle(G(p)-\langle G(p) \rangle)^2 \rangle}$.
This quantity is the standard deviation of $G(p)$ for a given $p$. We can check that $\chi_2$ also 
collapses well onto a single curve, 
\begin{equation}
\chi_2(p) \propto N^{\gamma_2/\overline{\nu}} f_2((p-p_d)N^{1/\overline{\nu}}).\label{susc2}
\end{equation} 
with $\gamma_2/\overline{\nu} = 1$ in Fig. \ref{fig1}(f). 
This result suggests that the PT is indeed discontinuous. 

Similar analyses are carried out in three and four dimensions.
In three dimensions, we obtain results similar to those of two dimensions but with different exponent values, 
i.e., $1/\overline{\nu}=0.70\pm 0.02$, $\gamma_1/\overline{\nu}=0.90\pm 0.02$, 
and $\gamma_2/\overline{\nu}=1.00\pm 0.02$. The data from different system sizes 
collapse well onto the scaling functions, which are shown in Fig.~\ref{fig2}. 
In four dimensions, we obtain similar results but with different exponent values, 
i.e., $1/\overline{\nu}=0.60\pm 0.02$, $\gamma_1/\overline{\nu}=0.92\pm 0.01$, 
and $\gamma_2/\overline{\nu}=1.00\pm 0.01$. The data from different system sizes 
collapse well onto the scaling functions as shown in Fig.~\ref{fig3}. 
These results verify the discontinuity of PT in three and four dimensions.

We also investigate the discontinuous PT in a different approach via the Smoluchowski 
equation, which describes the dynamics of cluster aggregations. 
In particular, we introduce an asymmetric  Smoluchowski equation in which the collision kernel 
is different depending on whether cluster is mobile \cite{dlca} as follows:
\begin{equation}
\frac{dn_s}{dp}=\sum_{i+j=s}\frac{k^{\prime}_ik_j}
{C^{\prime}(p)C(p)}n_i n_j-\frac{n_sk^{\prime}_s}{C^{\prime}(p)}-\frac{n_sk_s}{C(p)},\label{smolo}
\end{equation}
where $n_s \equiv N_s/N$ is the concentration of $s$-sized clusters, and $k^{\prime}_i/C^{\prime}$ and 
$k_j/C$ are the collision kernels of immobile and mobile clusters, respectively. 
$C^{\prime} \equiv \sum_{s}k^{\prime}_sn_s$ and $C \equiv \sum_{s}k_sn_s$ are the 
normalization factors. 
The first term on the right-hand side of Eq.~(\ref{smolo}) represents the 
aggregation of a mobile cluster of size $i$ and an immobile cluster of size $j$ with $s=i+j$. The
second term represents a mobile cluster of size $s$ merging with an immobile cluster of any size
including the largest cluster, in which $\sum_j k_jn_j = C(p)$ is used. The third term represents an
immobile cluster of size $s$ merging with a mobile cluster of any size including the largest size,
in which $\sum_j k_j^{\prime} n_j = C^{\prime}(p)$ is used. The summation runs 
only for finite clusters. Once an infinite-sized cluster is formed, the dynamics is terminated. 
We also do not need to consider finite clusters and infinite cluster separately as in sol-gel transitions 
\cite{ziff_1983}.
The collision kernel is determined by intuitive argument 
as the perimeter of clusters and thus $k^{\prime}_i/C^{\prime} \sim i^{\eta + 1- 1/d_f}$ 
for mobile and $k_i/C \sim i^{1-1/d_f}$ for immobile clusters, where $d_f$ is the fractal dimension 
of clusters \cite{ernst} and $\eta =-0.5$ for the Brownian case.

In two dimensions, using $d_f \approx 1.4$, it is estimated that $k_i^{\prime}\sim i^{-0.2}$ 
and $k_i\sim i^{0.3}$, which are in agreement with the numerical estimations 
of $k_i^{\prime}\sim i^{-0.2}$ and $k_i\sim i^{0.4}$ shown in Figs.~\ref{fig1}(g) and 1(h), respectively.  
In three dimensions, $d_f \approx 1.8$ \cite{three_fdim}, 
and thus $k^{\prime}_i \sim i^{-0.1}$ and $k_i \sim i^{0.4}$ are expected, which are again in 
agreement with the measured values $k^{\prime}_i \sim i^{-0.1}$ and $k_i \sim i^{0.4}$
as shown in Fig.~\ref{fig4}.  In four dimensions, $d_f \approx 2.0$ \cite{three_fdim}, and thus 
 $k^{\prime}_i \approx {\rm const}$ and $k_i \sim i^{0.5}$ are expected, which are again in 
agreement with the measured values $k^{\prime}_i \sim i^{0.0}$ and $k_i \sim i^{0.6}$ 
as shown in Fig.~\ref{fig5}.

Next, we investigate the growth of the giant component size $G$ by simulating the Smoluchowski equation 
numerically with the collision kernels we measured. Starting from $N$ monomers initially, 
numerical simulations are carried out as a function of $p$ for different system sizes. 
We plot the giant cluster size $G(p)$ and the susceptibilities $\chi_1$ and $\chi_2$ as a function of $p$ 
in scaling forms, and we find that the data of different system sizes collapse well onto a single curve with the 
critical exponents previous obtained as shown in Fig.\ref{fig6}. 

\subsection{Generalization of velocity scaling}

In this section, we study the PT of the DLCA model. Here, the scaling of the collision rate is generalized 
for computational simplicity by shifting the scaling with cluster size entirely into the scaling of 
the cluster velocity: $v_s \sim s^{\eta}$. It is then necessary to know whether the scaling exponent 
of the cluster velocity $\eta$ can be positively valued. 
Consider the motion of a small solid sphere introduced into a non-uniform electric field in air. 
This arrangement is readily realized in a corona discharge \cite{corona}, for example, between a grounded 
hollow cylinder and a thin wire along the cylinder's axis, when the wire is charged to a negative high voltage 
with respect to the cylinder. The corona current sustains drifting negative ions towards the cylinder walls. 
The sphere attracts the ions onto its surface by the image charge effect until the Coulomb repulsion 
by the accumulated ions prohibits it. The sphere is accelerated by the local electric field while its motion 
is resisted by the Stokes drag \cite{stokes}, reaching a terminal velocity that scales as the radius of the 
sphere. In 3-D the mean velocity scales as $v_s \sim s^{1/3}$, and the collision rate scales as $\sigma v_s \sim 
s$ for solid spheres. For fractal spheres, as in the DLCA model, $\eta$ can become positive. 

 To implement the effect of this velocity in simulations, we select an $s$-sized cluster  with a probability proportional to $s^{\eta}$, allow the cluster to diffuse to a nearest neighbor, 
 and make time pass by $\delta t=1/(\sum_s N_s s^{\eta})$.
If the cluster comes into contact with another cluster, the variable $p$ is advanced by $1/N$, regardless of the 
cluster size $s$.  As $\eta$ increases in positive region, the velocity of large clusters becomes large, 
so that they have higher 
probability of merging with another cluster. Thus, the growth rate of larger clusters is higher.  
This behavior was originally observed by Meakin {\it et al}.~\cite{meakin}. They argued that
the exponent of cluster size distribution increases as the velocity exponent increases.
In this case, the PT is continuous, because the giant cluster grows continuously. In contrast, when $\eta$ is negative, 
large-sized clusters are suppressed in growth, and their number is reduced. Instead, medium-sized clusters become 
abundant. As $p$ increases, such medium-sized clusters merge suddenly and create a giant cluster. 
Thus, the PT is discontinuous. Fig. \ref{fig7} shows the snapshots of clusters for different values of $\eta$ 
just before the percolation threshold. 
From these properties, one can guess that the transition type changes from discontinuous to continuous 
as $\eta$ increases across a certain value $\eta_c$. 

To determine the tricritical point $\eta_c$, we start from $\eta = -0.5$ and 
observe the change of the cluster size distribution by increasing $\eta$. 
In our previous study~\cite{precan}, it was shown that discontinuous (continuous) PT occurs
when $n_s(p_c^+) \sim s^{-\tau}$ satisfies $\tau < 2$ $(\tau > 2)$. 
We use this result to determine the tricritical point $\eta_c$.
Fig.~\ref{fig8} shows the cluster size distribution for several values of $\eta$ near $\eta_c$.
We estimate $\eta_c \approx 1.3$ in two dimensions [Fig.~\ref{fig8}(a)], 
$\approx 0.8$ in three dimensions [Fig.~\ref{fig8}(b)] and $\approx 0.4$ in four dimensions 
[Fig.~\ref{fig8}(c)] based on numerical data.

To confirm the continuity of PT in the region $\eta > \eta_c$, we perform finite-size analysis for $G(p)$.
The behavior of $G(p)$ with $\eta=1.5 > \eta_c\approx 1.3$ in two dimensions is plotted in Fig.~\ref{fig1}(i).
As we can observe in Fig.~\ref{fig1}(i), the crossing point of $G$ between two different sizes  
decreases as the system size increases, which means that the transition is continuous. But this tendency 
cannot be seen clearly in this figure. Thus we attempt a finite-size scaling analysis for continuous transition.
In Fig.~\ref{fig1}(j), the $G(p)$ of the different system sizes used in Fig.~\ref{fig1}(i) is collapsed
onto a single curve in the scaling form.
\begin{equation}
G_N(p) \propto N^{-\beta/\nu}f_0((p-p_c(\infty))N^{1/\nu}).\label{cont_fss}
\end{equation} 
A similar analysis is carried out in three dimensions, which is shown in Fig.~\ref{fig9}(a).
In this analysis, we use the system of $\eta=1.0 > \eta_c \approx 0.8$. 
In the inset, we use the scaling form given by Eq.~(\ref{cont_fss}).
To verify the continuous transition in an alternative way, we use the Smoluchowski equation.
Similar to what we found in the previous subsection, we obtain $k^{\prime}_i/C^{\prime}\approx i^{1.35}$ 
and $k_i/C \approx i^{0.20}$ at $p_d$ in Fig.~\ref{fig4}. 
Fig.~\ref{fig9}(b) shows the simulation result of the Smoluchowski equation for various system 
sizes. In the inset, we use the scaling form given by Eq.~(\ref{cont_fss}) to verify continuity and find 
that data are well collapsed on a single curve. 
These results confirm that the transition is indeed continuous in the region $\eta > \eta_c \approx 0.8$
in three dimensions. 

\begin{figure}
\includegraphics[width=0.9\linewidth]{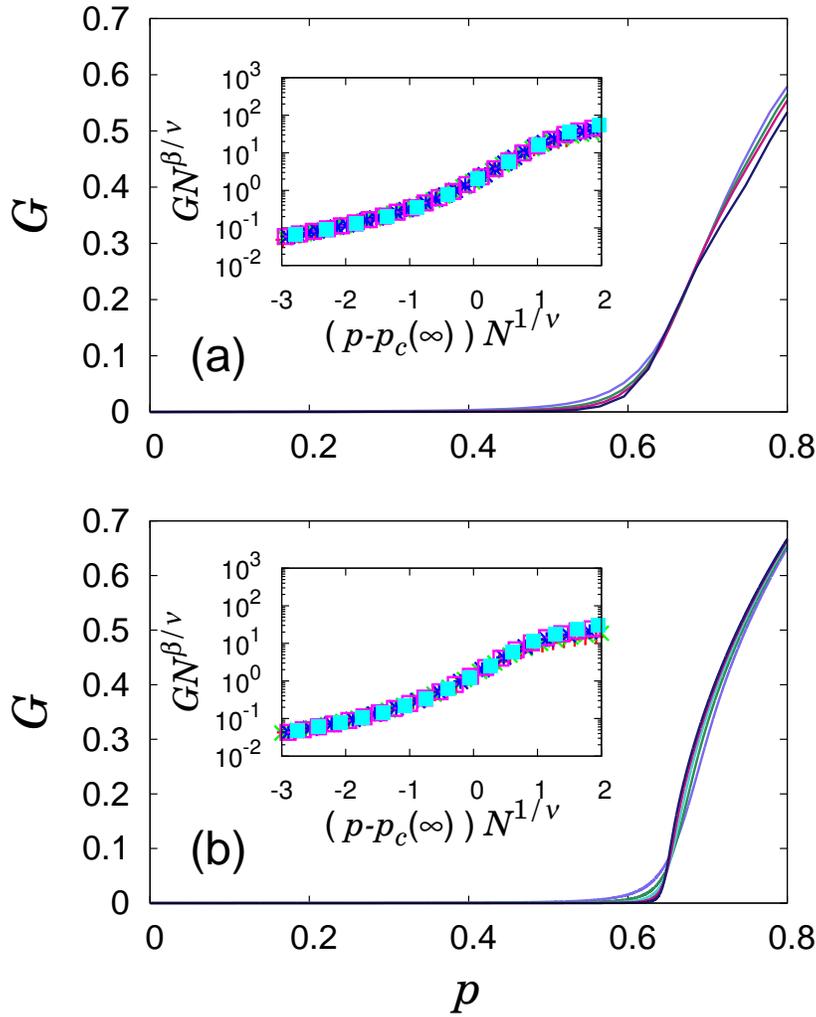}
\caption{(Color online) (a) Simulation result of DLCA with velocity exponent 
$\eta=0.8 > \eta_c \approx 0.4$ in four dimensions.
(b) Numerical simulation of the Smoluchowski equation using the collision kernel $k^{\prime}_i/C^{\prime} \sim i^{1.30}$ and 
$k_i/C \sim i^{0.40}$. 
Inset in (a): Data collapse of the data used in (a). We perform simulations with linear sizes  $L/10 = 6, 8, 10, 12$, and $16$ with particle density $\rho = 0.0004$. Data of different system sizes collapse well onto a single curve predicted theoretically for a continuous transition in which we use $p_c(\infty)=0.56$, $1/\nu=0.22$, and $\beta/\nu=0.43$. 
The behavior of the data collapse indicates that the transition is indeed continuous. 
Inset in (b): Data collapse of the data used in (b). Sizes of the systems are $N/10^4 = 8, 32, 128, 512$ and $2048$. We use $p_c(\infty)=0.64$, $1/\nu=0.25$, and $\beta/\nu=0.3$ for data collapse. Again the data-collapse behavior indicates a continuous PT.}
\label{fig10}
\end{figure}

A similar analysis used for three dimensions is also applied for four dimensions, which is shown in Fig.~\ref{fig10}.
In Fig.~\ref{fig10}(a), the $G(p)$ of various system sizes are plotted and these data are well collapsed
onto a single curve if we use the previous scaling form given by Eq.~(\ref{cont_fss}), which is shown in the inset.
Second, we take $k^{\prime}_i \sim i^{1.30}$ and $k_i \sim i^{0.40}$ from Fig.~\ref{fig5} to simulate
the Smoluchowski equation in Fig.~\ref{fig10}(b). In the inset, the data are well collapsed in the
previous scaling form Eq.~\ref{cont_fss}. These results again confirm the continuous transition in the region 
$\eta > \eta_c \approx 0.4$ in four dimensions.

\begin{figure*}
\includegraphics[width=0.9\linewidth]{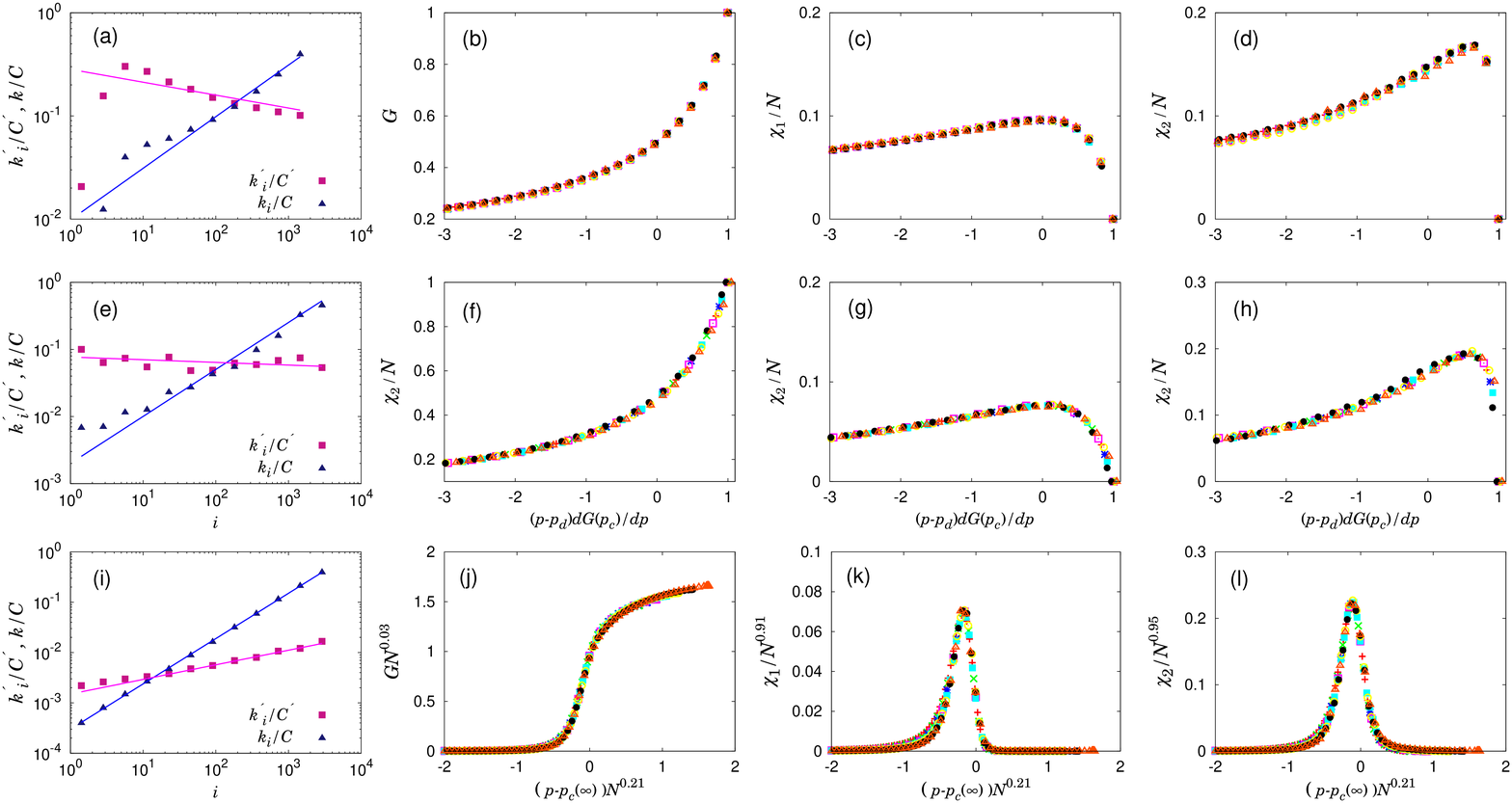}
\caption{(Color online) (a)-(d) Plot of the RLCA model in the Brownian process in two dimensions.  (a) Plot of $k^{\prime}_i/C^{\prime}$ and $k_i/C$ vs $i$ at $p_d$. $N = 2000$ and $L = 100$ are used in the simulations.
The two collision kernels are estimated as $k^{\prime}_i/C^{\prime} \sim i^{-0.12 \pm 0.06}$ and $k_i/C \sim i^{0.50 \pm 0.03}$. (b)-(d) Plots of finite-size scaling behaviors of (b) $G$, (c) $\chi_1$, and (d) $\chi_2$.
simulations of the Smoluchowski equation are performed using $N/10^4 = 16, 32, 64, 128, 256, 512, 1024$, and $2048$. Data are well collapsed onto the scaling curves for the discontinuous transition with 
$dG(p_c)/dp \sim N$ in two dimensions, which indicates that the PT of the RLCA model in the Brownian process is discontinuous. 
Plots (e)-(h) correspond to plots (a)-(d) but in three dimensions. For (e), $N = 4000$ and $L = 100$ are used. 
The collision kernels are estimated as $k^{\prime}_i/C^{\prime} \sim i^{-0.04 \pm 0.03}$ and $k_i/C \sim i^{0.70 \pm 0.04}$.
In (f)-(h), the system sizes are $N/10^4 = 16, 32, 64, 128, 256, 512, 1024$ and $2048$. Data collapse is well behaved using the formulas for the discontinuous transition with $dG(p_c)/dp \sim N^{0.75}$.  
Plots (i)-(l) correspond to the plots (a)-(d) but in four dimensions. For (i), $N = 5184$ and $L = 60$ are used. 
The collision kernels are estimated as $k^{\prime}_i/C^{\prime} \sim i^{0.29 \pm 0.01}$ and $k_i/C \sim i^{0.90 \pm 0.00}$. 
For (j)-(l), the system sizes are $N/10^4 = 16, 32, 64, 128, 256, 512, 1024$ and $2048$. Data are well collapsed on the scaling formulas for the continuous transition. It is estimated that $p_c(\infty) = 0.95$, $1/\nu = 0.21$, and $\beta/\nu = 0.05$.}\label{fig11}
\end{figure*}

\section{Reaction-limited cluster aggregation model}

Here, we perform similar studies for the reaction-limited cluster aggregation (RLCA) model in the Brownian process in two, three, and four dimensions. In this model, two adjoining clusters merge irreversibly with probability $r$, but with the remaining with probability $1-r$, they can move independently.  As $r$ goes to 0, a cluster can penetrate inside the area between branches of another cluster, becoming trapped and irreversibly stuck within it. As a result, the resulting cluster becomes less ramified, and its fractal dimension is increased \cite{rlca}. Here we use $r = 10^{-3}$, in which the dynamics of cluster aggregations of the RLCA observed is different from that of the DLCA. 

To obtain the giant cluster size $G$, we measure the collision kernels $k^{\prime}_i/C^{\prime}$ and $k_i/C$ at $p_d$ and simulate the Smoluchowski equation using these collision kernels. Monte Carlo simulations of the RLCA model take extremely long computation times for us to understand the finite-size scaling behavior of the PT. Thus, we measure the collision kernels in two, three, and four dimensions, and then investigate the finite-size scaling behavior of numerical data of the Smoluchowski equations. The measured collision kernels are shown in Fig.~\ref{fig11}. The collision kernel may be written in a power-law form, $k^{\prime}_i k_j \sim i^{\omega^{\prime}}j^{\omega}$. Then the exponents ($\omega^{\prime}$, $\omega$) are estimated as ($-0.12 \pm 0.06, 0.50 \pm 0,03$), ($-0.04 \pm 0.03, 0.70 \pm 0.04$) and ($0.29 \pm 0.10, 0.90 \pm 0.00$) in two, three and four dimensions, respectively. It is noteworthy that in the conventional Smoluchowski equation, clusters are immobile, and thus the collision kernel is symmetric as $k_ik_j$. If $k_ik_j \sim (ij)^{\zeta}$, then for $\zeta > 0.5$, the percolation transition is continuous. In this case, the cluster-size distribution follows a power law at the critical point as $n_s \sim s^{-\tau}$, 
where $\tau > 2$~\cite{ziff_1983}. However, for the asymmetric case above, the criterion for continuous transitions is
not known specifically to our knowledge. We find that the cluster size distribution for the asymmetric Smoluchowski 
equation for the RLCA model with the numerically estimated kernels at the transition point follows a power law,
$n_s \sim s^{-0.60}$ in two dimensions, $n_s \sim s^{-1.25}$ in three dimensions up to finite-size cutoffs, 
and $n_s \sim s^{-2.05}$ in four dimensions. 
Thus, the percolation transition for four dimensions can be expected to be continuous. 
We also remark that the collision kernel for the RLCA does not agree well with the one obtained from the formulas $k^{\prime}_i \sim i^{1-1/d_f+\eta}$ and $k_i \sim i^{1-1/d_f}$. If we use $d_f=1.7$, $d_f=2.0$, and $d_f=2.4$ for two, three and four dimensions, respectively, and $\eta=-0.5$ \cite{rlca2}. This difference is because merging of 
two clusters does not occur at their perimeters in the RLCA process. Using these obtained collision kernels, we perform numerical simulations of the Smoluchowski equation, and we find that the giant cluster size $G$ and the susceptibilities $\chi_1$ and $\chi_2$ behave following the scaling functions  
for the discontinuous PT in Eqs.(\ref{g_scaling}-\ref{susc2}) in two and three dimensions. However, in four dimensions, the obtained data do not collapse onto the scaling functions of the discontinuous transition, but rather collapse onto Eq.~(\ref{cont_fss}), valid for continuous transitions. 
This different behavior is caused by the large exponent values of the collision kernels. 

\section{Summary}

In this paper, we extended the previous study of discontinuous PT of the diffusion-limited cluster aggregation (DLCA) model to three and four dimensions. We showed that the discontinuous PT also occurs even in three and four dimensions for Brownian motion. In this case, the discontinuous PT is caused by the natural suppression effect of Brownian motion to the growth of large clusters.
Moreover, we studied PT for the DLCA model with general velocity $v_s \sim s^{\eta}$ for various values of $\eta$, where $s$ is the cluster size. As $\eta$ increases, the suppression effect becomes weak, so that there exists a tricritical point $\eta_c$, across which the PT type changes from 
discontinuous to continuous. Finally, we briefly studied the PT for the reaction-limited cluster aggregation (RLCA) model in Brownian motion in two, three and four dimensions. By simulating the Smoluchowski equation with the obtained collision kernels, we find that the PT is discontinuous in two and three dimensions but continuous in four dimensions. 
In this work, $p_c \rightarrow 1$ for the cases of discontinuous transitions, otherwise $p_c < 1$ in the cases of continuous transitions. Conclusively, we expect that the discontinuous PT can be observed in many modified DLCA models owing to the suppression effect of Brownian motion. 

\ack

This study was supported by NRF grants funded by MEST (Grant No. 2010-0015066), the Brain pool program (YWK), and the Seoul Science Foundation and the Global Frontier program (YSC).


\section*{References}
\begin{thebibliography}{99}
\bibitem{stauffer} Stauffer D and Aharony A, 1994 {\it Introduction to Percolation Theory} (Taylor \& Francis, London)
\bibitem{er} Erd\H{o}s P and R\'enyi A, 1960 Publ. Math. Inst. Hung. Acad. Sci. {\bf 5,} 17
\bibitem{ap} Achlioptas D, D'Souza R M and Spencer J, 2009 {\it Science} {\bf 323}, 1453
\bibitem{friedman} Friedman E J and Landsberg A S, 2009 Phys. Rev. Lett. {\bf 103,} 255701
\bibitem{fortunato} Radicchi F and Fortunato S, 2010 Phys. Rev. E {\bf 81,} 036110
\bibitem{dorogovtsev} da Costa R A, Dorogovtsev S N, Goltsev A V and Mendes J F F, 2010
Phys. Rev. Lett. {\bf 105,} 255701
\bibitem{hklee}  Lee H K, Kim B J and Park H, 2011 Phys. Rev. E {\bf 84,} 020101(R)
\bibitem{grassberger} Grassberger P, Christensen C, Bizhani G, Son S-W and Paczuski M, 2011
Phys. Rev. Lett. {\bf 106,} 225701
\bibitem{science} Riordan O and Warnke L, 2011 Science {\bf 333,} 322
\bibitem{suppression} Cho Y S and Kahng B, 2011 Phys. Rev. Lett. {\bf 107}, 275703
\bibitem{souza} D'Souza R M and Mitzenmacher M, 2010 Phys. Rev. Lett. {\bf 106,} 115701
\bibitem{hermann} Schrenk K J, Ara\'ujo N A M and Herrmann H J, 2011 Phys. Rev. E {\bf 84,} 041136
\bibitem{choi} Choi W, Yook S H and Kim Y, 2011 Phys. Rev. E {\bf 84,} 020102
\bibitem{ziff} Boettcher S, Singh V and Ziff R M, 2012 Nat. Commun. {\bf 3}, 787
\bibitem{dlca} Cho Y S and Kahng B, 2011 Phys. Rev. E {\bf 84,} 050102(R)
\bibitem{meakin} Meakin P, 1983 Phys. Rev. Lett. {\bf 51,} 1119; Meakin P, Vicsek T and Family F, 1985 Phys. Rev. B {\bf 31,} 564 
\bibitem{kolb} Kolb M, Botet R and Jullien R, 1983 Phys. Rev. Lett. {\bf 51,} 1123; Kolb M and Jullien R, 1984 J. Phys. (France) Lett. {\bf 45}, L977
\bibitem{kim} Kim Y W, Lee H and Belony, Jr P, 2006 Rev. Scientific Inst. {\bf 77,} 10F115 
\bibitem{fss} Cho Y S, Kim S-W, Noh J D, Kahng B and Kim D, 2010 Phys. Rev. E {\bf 82,} 042102
\bibitem{ziff_1983} Ziff R M,  Hendriks E M, and Ernst M H, 1983 J. Phys. A 16, 2293
\bibitem{ernst} Ernst M H, Hendriks E M, and Leyvraz F, 1984 J. Phys. A {\bf 17}, 2137
\bibitem{three_fdim} Jullien R, Kolb M and Botet R, 1984 J. Phys. (Paris), Lett. {\bf 45,} L211
\bibitem{corona} Goldman M, Goldman A and Sigmond R S, 1985 Pure and Appl. Chem. {\bf 57}, 1353  
\bibitem{stokes} Kundu P K and Cohen I M, 2004 {\it Fluid Mechanics 3rd edition} (Elsevier Academic Press, San Diego) 
\bibitem{precan} Cho Y S, Kahng B and Kim D, 2010 Phys. Rev. E {\bf 81,} 030103(R)
\bibitem{rlca} Kolb M and Jullien R, 1984 J. Phys. (Paris), Lett. {\bf 45,} L977
\bibitem{rlca2} Kolb M, 1986 J. Phys. A {\bf 19,} L263


\end{thebibliography}
\end{document}